\documentclass[english]{article}
\usepackage[T1]{fontenc}
\usepackage[latin9]{inputenc}
\usepackage{amssymb}
\usepackage{esint}

\makeatletter
\newcommand{\lyxaddress}[1]{
\par {\raggedright #1
\vspace{1.4em}
\noindent\par}
}

\makeatother

\usepackage{babel}

\begin{document}

\title{Entropy of the Kerr-Sen Black Hole}

\author{Alexis Larrañaga}

\maketitle

\lyxaddress{National University of Colombia. National Astronomical Observatory.}
\begin{abstract}
We study the entropy of Kerr-Sen black hole of heterotic string theory
beyond semiclassical approximations. Applying the properties of exact
differentials for three variables to the first law thermodynamics
we derive the corrections to the entropy of the black hole. The leading
(logarithmic) and non leading corrections to the area law are obtained.
\end{abstract}

\section{Introduction}

When studying black hole evaporation by Hawking radiation using the
quantum tunneling approach, a semiclassical treatment is used to study
changes in thermodynamical quantities. The quantum corrections to
the Hawking temperature and the Bekenstein- Hawking area law have
been studied for the Schwarzschild, Kerr and Kerr-Newman black holes
\cite{otros,mann} as well as BTZ black holes \cite{akbar}. 

It has been realized that the low-energy effective field theory describing
string theory contains black hole solutions which can have properties
which are qualitatively different from those that appear in ordinary
Einstein gravity. Here we will analyze the quantum corrections to
the entropy of the Kerr-Sen black hole, which is an exact classical
solution in the low-energy effective heterotic string theory with
a finite amount of charge and angular momentum. To obtain the quantum
corrections we use the criterion for exactness of differential of
black hole entropy, from the first law of thermodynamics with three
parameters. We find that the leading correction term is logarithmic,
while the other terms involve ascending powers of inverse of the area. 

In the quantum tunneling approach, when a particle with positive energy
crosses the horizon and tunnels out, it escapes to infinity and appear
as Hawking radiation. Meanwhile, when a particle with negative energy
tunnels inwards it is absorbed by the black hole and as a result the
mass of the black hole decreases. Therefore, the essence of the quantum
tunneling argument for Hawking radiation is the calculation of the
imaginary part of the action. If we consider the action $\mathcal{I}\left(r,t\right)$
and make an expansion in powers of $\hbar$ we obtain

\begin{eqnarray}
\mathcal{I}\left(r,t\right) & = & \mathcal{I}_{0}\left(r,t\right)+\hbar\mathcal{I}_{1}\left(r,t\right)+\hbar^{2}\mathcal{I}_{2}\left(r,t\right)+...\\
 & = & \mathcal{I}_{0}\left(r,t\right)+\sum_{i}\hbar^{i}\mathcal{I}_{i}\left(r,t\right),\end{eqnarray}

where $\mathcal{I}_{0}$ gives the semiclassical value and the terms
from $O\left(\hbar\right)$ onwards are treated as quantum corrections.
The work of Banerjee and Majhi \cite{expansion} shown that the correction
terms $\mathcal{I}_{i}$ are proportional to the semiclassical contribution
$\mathcal{I}_{0}$. Since $\mathcal{I}_{0}$ has the dimension of
$\hbar$, the proportionality constants should have the dimension
of inverse of $\hbar$. In natural units $\left(G=c=k_{B}=1\right)$,
the Planck constant is of the order of square of the Planck Mass.
Therfore, from dimensional analysis the proportionality constants
have the dimension of $M^{-2i}$ where $M$ is the mass of black hole,
and the series expansion becomes

\begin{eqnarray}
\mathcal{I}\left(r,t\right) & = & \mathcal{I}_{0}\left(r,t\right)+\sum_{i}\beta_{i}\frac{\hbar^{i}}{M^{2i}}\mathcal{I}_{0}\left(r,t\right)\\
 &  & \mathcal{I}_{0}\left(r,t\right)\left(1+\sum_{i}\beta_{i}\frac{\hbar^{i}}{M^{2i}}\right),\end{eqnarray}
where $\beta_{i}$'s are dimensionless constant parameters. If the
black hole has other macroscopic parameters such as angular momentum
and electric charge, one can express this expansion in terms of the
area of the event horizon, i.e. using the horizon radius $r_{H}$
and the angular momentum $a$, as done in \cite{otros} and \cite{akbar},

\begin{eqnarray}
\mathcal{I}\left(r,t\right) & = & \mathcal{I}_{0}\left(r,t\right)\left(1+\sum_{i}\beta_{i}\frac{\hbar^{i}}{\left(r_{H}^{2}+a^{2}\right)^{i}}\right).\label{eq:actionexpansion}\end{eqnarray}

This expansion will be used later to calculate the quantum corrections
to the entropy of the Kerr-Sen black hole.

\section{Entropy as an Exact Differential}

In order to perform the quantum corrections to the entropy of the
black hole we will follow the analysis of \cite{otros} and \cite{akbar}. The first
law of thermodynamics for charged and rotating black holes is 

\begin{equation}
dM=TdS+\Omega dJ+\Phi dQ,\end{equation}
where the parameters $M,J$ and $Q$ are the mass, angular momentum
and charge of the black hole, respectively, while $T,S,\Omega$ and
$\Phi$ are the temperature, entropy, angular velocity and electrostatic
potential, respectively. This equation can be rewritten as

\begin{equation}
dS\left(M,J,Q\right)=\frac{1}{T}dM-\frac{\Omega}{T}dJ-\frac{\Phi}{T}dQ,\end{equation}
from which one can infer that, in order for $dS$ to be an exact differential,
the thermodynamical quantities must satisfy

\begin{eqnarray}
\frac{\partial}{\partial J}\left(\frac{1}{T}\right) & = & \frac{\partial}{\partial M}\left(-\frac{\Omega}{T}\right)\label{eq:cond1}\\
\frac{\partial}{\partial Q}\left(\frac{1}{T}\right) & = & \frac{\partial}{\partial M}\left(-\frac{\Phi}{T}\right)\label{eq:cond2}\\
\frac{\partial}{\partial Q}\left(-\frac{\Omega}{T}\right) & = & \frac{\partial}{\partial J}\left(-\frac{\Phi}{T}\right).\label{eq:cond3}\end{eqnarray}

If $dS$ is an exact differential, we can write the entropy $S(M,J,Q)$
in the integral form

\begin{eqnarray}
S\left(M,J,Q\right) & = & \int\frac{1}{T}dM-\int\frac{\Omega}{T}dJ-\int\frac{\Phi}{T}dQ-\int\left(\frac{\partial}{\partial J}\left(\int\frac{1}{T}dM\right)\right)dJ\nonumber \\
 &  & -\int\left(\frac{\partial}{\partial Q}\left(\int\frac{1}{T}dM\right)\right)dQ+\int\left(\frac{\partial}{\partial Q}\left(\int\frac{\Omega}{T}dJ\right)\right)dQ\nonumber \\
 &  & +\int\left(\frac{\partial}{\partial Q}\left(\int\left(\frac{\partial}{\partial J}\left(\int\frac{1}{T}dM\right)\right)dJ\right)\right)dQ.\label{eq:integralentropy}\end{eqnarray}

\section{Standard Entropy of the Kerr-Sen black hole}

Sen \cite{Sen,Senotro} was able to find a charged, stationary, axially
symmetric solution of the field equations by using target space duality,
applied to the classical Kerr solution. The line element of this solution
can be written, in generalized Boyer-Linquist coordinates, as

\begin{eqnarray}
ds^{2} & = & -\left(1-\frac{2Mr}{\rho^{2}}\right)dt^{2}+\rho^{2}\left(\frac{dr^{2}}{\Delta}+d\theta^{2}\right)-\frac{4Mra\sin^{2}\theta}{\rho^{2}}dtd\varphi\nonumber \\
 &  & +\left(r\left(r+r_{\alpha}\right)+a^{2}+\frac{2Mra^{2}\sin^{2}\theta}{\rho^{2}}\right)\sin^{2}\theta d\varphi^{2},\label{eq:kerrsen}\end{eqnarray}
where 

\begin{eqnarray}
\Delta & = & r\left(r+r_{\alpha}\right)-2Mr+a^{2}\\
\rho & = & r\left(r+r_{\alpha}\right)+a^{2}\cos^{2}\theta.\end{eqnarray}

Here $M$ is the mass of the black hole, $a=\frac{J}{M}$ is the specific
angular momentum of the black hole and the electric charge is given
by

\begin{equation}
r_{\alpha}=\frac{Q^{2}}{M}.\end{equation}
Note that in the particular case of a static black hole, i.e. $a=0$,
the metric (\ref{eq:kerrsen}) coincides with the GMGHS solution \cite{gmghs}
while in the particular case $r_{\alpha}=0$ reconstructs the Kerr
solution.

The Kerr-Sen space has a spherical event horizon, which is the biggest
root of the equation $\Delta=0$ and is given by

\[
r_{H}=\frac{2M-r_{\alpha}+\sqrt{\left(2M-r_{\alpha}\right)^{2}-4a^{2}}}{2}\]

or in terms of the black hole parameters $M,Q$ and $J$,

\begin{equation}
r_{H}=M-\frac{Q^{2}}{2M}+\sqrt{\left(M-\frac{Q^{2}}{2M}\right)^{2}-\frac{J^{2}}{M^{2}}}.\label{eq:horizon}\end{equation}
The area of the event horizon is given by

\begin{equation}
A=4\pi\left(r_{H}^{2}+a^{2}\right)=8\pi M\left(M-\frac{Q^{2}}{2M}+\sqrt{\left(M-\frac{Q^{2}}{2M}\right)^{2}-\frac{J^{2}}{M^{2}}}\right).\label{eq:area}\end{equation}

Equation (\ref{eq:horizon}) tell us that the horizon disappears unless

\[
\left|J\right|\leq M^{2}-\frac{Q^{2}}{2},\]
therefore, the extremal black hole, $\left|J\right|=M^{2}-\frac{Q^{2}}{2}$,
has $A=8\pi\left|J\right|$. The angular velocity at the horizon is
given by

\begin{equation}
\Omega=\frac{J}{2M^{2}}\frac{1}{M-\frac{Q^{2}}{2M}+\sqrt{\left(M-\frac{Q^{2}}{2M}\right)^{2}-\frac{J^{2}}{M^{2}}}}\end{equation}
 while the Hawking temperature is

\begin{equation}
T_{H}=\frac{\kappa\hbar}{2\pi}=\frac{\hbar\sqrt{\left(2M^{2}-Q^{2}\right)^{2}-4J^{2}}}{4\pi M\left(2M^{2}-Q^{2}+\sqrt{\left(2M^{2}-Q^{2}\right)^{2}-4J^{2}}\right)}.\end{equation}

One can easily check that thermodynamical quantities for the Kerr-Sen
black hole satisfy 

\begin{eqnarray}
\frac{\partial}{\partial J}\left(\int\frac{1}{T_{H}}dM\right) & = & -\frac{\Omega}{T_{H}}\label{eq:cond1a}\\
\frac{\partial}{\partial Q}\left(\int\frac{1}{T_{H}}dM\right) & = & -\frac{\Phi}{T_{H}}\label{eq:cond2a}\\
\frac{\partial}{\partial Q}\left(\int-\frac{\Omega}{T_{H}}dJ\right) & = & -\frac{\Phi}{T_{H}}.\label{eq:cond3a}\end{eqnarray}
Under this conditions, the integral form of the entropy (\ref{eq:integralentropy})
reduces to 

\begin{equation}
S_{0}\left(M,J,Q\right)=\int\frac{1}{T_{H}}dM.\end{equation}
For the Kerr-Sen black hole this gives,

\begin{equation}
S_{0}\left(M,J,Q\right)=\frac{4\pi}{\hbar}\int\frac{M\left(2M^{2}-Q^{2}+\sqrt{\left(2M^{2}-Q^{2}\right)^{2}-4J^{2}}\right)}{\sqrt{\left(2M^{2}-Q^{2}\right)^{2}-4J^{2}}}dM\end{equation}

\begin{equation}
S_{0}\left(M,J,Q\right)=\frac{\pi}{\hbar}\left(2M^{2}-Q^{2}+\sqrt{\left(2M^{2}-Q^{2}\right)^{2}-4J^{2}}\right),\end{equation}
that corresponds to the standard black hole entropy 

\begin{equation}
S_{0}\left(M,J,Q\right)=\frac{A}{4\hbar}=\frac{\pi\left(r_{H}^{2}+a^{2}\right)}{\hbar}.\label{eq:standardentropy}\end{equation}

\section{Quantum Correction of the Entropy}

When considering the expansion for the action (\ref{eq:actionexpansion}),
it affects the Hawking temperature by introducing some correction
terms \cite{otros,akbar,expansion}. Therefore the temperature is now given
by

\begin{eqnarray}
T & = & T_{H}\left(1+\sum_{i}\beta_{i}\frac{\hbar^{i}}{\left(r_{H}^{2}+a^{2}\right)^{i}}\right)^{-1},\label{eq:newtemperature}\end{eqnarray}
where $T_{H}$ is the standard Hawking temperature and the terms with
$\beta_{i}$ are quantum correction terms to the temperature. It is
not difficult to verify that the conditions to make $dS$ an exact
differential are satisfied when considering the new for of the temperature.
Therefore, the entropy with correction terms is given by

\begin{equation}
S\left(M,J,Q\right)=\int\frac{1}{T}dM=\int\frac{1}{T_{H}}\left(1+\sum_{i}\beta_{i}\frac{\hbar^{i}}{\left(r_{H}^{2}+a^{2}\right)^{i}}\right)dM\end{equation}
or 

\begin{equation}
S\left(M,J,Q\right)=\int\frac{1}{T_{H}}dM+\int\frac{\beta_{1}}{T_{H}}\frac{\hbar}{\left(r_{H}^{2}+a^{2}\right)}dM+\int\frac{\beta_{2}}{T_{H}}\frac{\hbar^{2}}{\left(r_{H}^{2}+a^{2}\right)^{2}}dM+...\end{equation}
This equation can be written as 

\begin{equation}
S\left(M,J,Q\right)=S_{0}+S_{1}+S_{2}+....,\end{equation}
where $S_{0}$ is the standard entropy given by equation (\ref{eq:standardentropy})
and $S_{1},S_{2},...$ are quantum corrections. The first of these
terms is

\begin{eqnarray}
S_{1} & = & \beta_{1}\hbar\int\frac{1}{T_{H}\left(r_{H}^{2}+a^{2}\right)}dM\\
 & = & 4\pi\beta_{1}\int\frac{M}{\sqrt{\left(2M^{2}-Q^{2}\right)^{2}-4J^{2}}}dM.\end{eqnarray}
Solving the integral, we obtain

\begin{equation}
S_{1}=\pi\beta_{1}\ln\left|2\left(2M^{2}-Q^{2}+\sqrt{\left(2M^{2}-Q^{2}\right)^{2}-4J^{2}}\right)\right|,\end{equation}
that can be written as

\begin{equation}
S_{1}=\pi\beta_{1}\ln\left|2\left(r_{H}^{2}+a^{2}\right)\right|.\end{equation}
The following terms can be written, in general, as

\begin{equation}
S_{j}=\beta_{j}\hbar^{j}\int\frac{1}{T_{H}\left(r_{H}^{2}+a^{2}\right)^{j}}dM\end{equation}
\begin{equation}
S_{j}=4\pi\beta_{j}\hbar^{j-1}\int\frac{MdM}{\left(2M^{2}-Q^{2}+\sqrt{\left(2M^{2}-Q^{2}\right)^{2}-4J^{2}}\right)^{j-1}\sqrt{\left(2M^{2}-Q^{2}\right)^{2}-4J^{2}}}.\end{equation}
By calculating the integral, we obtain

\begin{equation}
S_{j}=\frac{\pi\beta_{j}\hbar^{j-1}}{1-j}\left(2M^{2}-Q^{2}+\sqrt{\left(2M^{2}-Q^{2}\right)^{2}-4J^{2}}\right)^{1-j}\end{equation}

or 

\begin{equation}
S_{j}=\frac{\pi\beta_{j}\hbar^{j-1}}{1-j}\left(r_{H}^{2}+a^{2}\right)^{1-j}\end{equation}
for $j>1$. Therefore, the entropy with the quantum corrections is

\begin{eqnarray}
S\left(M,J,Q\right) & = & \frac{\pi\left(r_{H}^{2}+a^{2}\right)}{\hbar}+\pi\beta_{1}\ln\left|2\left(r_{H}^{2}+a^{2}\right)\right|\nonumber \\
 &  & +\sum_{j>1}\frac{\pi\beta_{j}\hbar^{j-1}}{1-j}\left(r_{H}^{2}+a^{2}\right)^{1-j}.\end{eqnarray}

If we put charge, $Q=0,$ we recover the corrections for the case
of the Kerr black black hole found in \cite{expansion}. On the other
hand, if the angular momentum is also put equal to zero we obtain
the entropy for the Schwarzschild black hole $\left(a=Q=0\right)$.
Finally, if only the angular momentum vanishes (i.e. $a=0$), we get
corrections for the GMGHS black hole \cite{larranaga2}.

Using equation (\ref{eq:area}), and doing a re-definition of the
$\beta_{i}$, we can write the entropy in terms of the area of the
horizon as

\begin{eqnarray}
S\left(M,J,Q\right) & = & \frac{A}{4\hbar}+\pi\beta_{1}\ln\left|A\right|+\sum_{j>1}\frac{\pi\beta_{j}\hbar^{j-1}}{1-j}\left(\frac{A}{4\pi}\right)^{1-j}.\end{eqnarray}

The first term in this expansion is the usual semiclassical entropy
while the second term is the logarithmic correction found earlier
for other geometries \cite{kumar,larranaga} and using different methods.
The value of the coefficients $\beta_{i}$ can be evalated using other
approaches, such as the entanglement entropy calculation. Finally
note that the third term in the expansion is an inverse of area term
similar to the one obtained by S. K. Modak \cite{kumar} for the rotating
BTZ black hole, for the charged BTZ black hole \cite{larranaga} and
also in the general cases studied in \cite{otros} and \cite{akbar}.

\section{Conclusion}

As is well known, the Hawking evaporation process can be understood
as a consequence of quantum tunneling in which some particles cross
the event horizon. The positive energy particles tunnel out of the
event horizon, whereas, the negative energy particles tunnel in, resulting
in black hole evaporation. Using this analysis we have studied the
quantum corrections to the entropy for Kerr-Sen black hole. With the
help of the conditions for exactness of differential of entropy we
obtain a power series for entropy. The first term is the semiclassical
value, while the leading correction term is logarithmic as has been
found using other methods\cite{kumar,larranaga}. The other terms
involve ascending powers of the inverse of the area. If the angular
momentum become zero, we obtain results for the GMGHS black hole,
in which case the power series involve just mass and electric charge\cite{larranaga2}.
This analysis shows that the quantum corrections to entropy obtained before \cite{otros,akbar}, also hold for the black holes of string theory studied here. 

\emph{Acknowledgements}. This work was supported by the Universidad
Nacional de Colombia. Project Code 2010100.

\end{document}